\documentclass{article}

\usepackage[final, nonatbib]{neurips_2023_ml4ps}

\usepackage[utf8]{inputenc} 
\usepackage[T1]{fontenc}    
\usepackage{hyperref}       
\usepackage{url}            
\usepackage{booktabs}       
\usepackage{amsfonts}       
\usepackage{nicefrac}       
\usepackage{microtype}      
\usepackage{xcolor}         
\usepackage[numbers]{natbib}

\usepackage{graphicx,pstricks}
\usepackage{graphics}
\usepackage{epsfig}
\usepackage{amssymb}
\usepackage{amsmath, amsthm}
\usepackage{amsfonts}
\usepackage{algorithm}
\usepackage{tabu}
\usepackage{multirow}
\usepackage{subfig}
\usepackage{subcaption}
\usepackage{caption}
\usepackage{changepage}
\usepackage{algpseudocode}
\usepackage{xcolor}

\title{Data-Driven Autoencoder Numerical Solver with Uncertainty Quantification for Fast Physical Simulations}

\author{
  Christophe Bonneville\thanks{Now at Sandia National Laboratories} \\
  Department of Civil \& Environmental Engineering\\
  Cornell University\\
  Ithaca, NY 14850 \\
  \texttt{cpb97@cornell.edu}
  \AND
  Youngsoo Choi\\
  Center for Applied Scientific Computing\\
  Lawrence Livermore National Laboratory\\
  Livermore, CA, 94550\\
  \texttt{choi15@llnl.gov}
  \AND
  Debojyoti Ghosh\\
  Center for Applied Scientific Computing\\
  Lawrence Livermore National Laboratory\\
  Livermore, CA, 94550\\
  \texttt{ghosh5@llnl.gov}\\
  \AND
  Jonathan L. Belof\\
  Materials Science Division\\
  Physical and Life Science Directorate\\
  Lawrence Livermore National Laboratory\\
  Livermore, CA, 94550\\
  \texttt{belof1@llnl.gov}}
    
\begin{document}

\maketitle

\begin{abstract}
Traditional partial differential equation (PDE) solvers can be computationally expensive, which motivates the development of faster methods, such as reduced-order-models (ROMs). We present GPLaSDI, a hybrid deep-learning and Bayesian ROM. GPLaSDI trains an autoencoder on full-order-model (FOM) data and simultaneously learns simpler equations governing the latent space. These equations are interpolated with Gaussian Processes, allowing for uncertainty quantification and active learning, even with limited access to the FOM solver. Our framework is able to achieve up to 100,000 times speed-up and less than 7$\%$ relative error on fluid mechanics problems.
\end{abstract}

\section{Introduction}

Over the last few decades, advances in numerical simulation methods and cheaper computational power have expanded the use of simulation across various fields, such as engineering design, digital twins, and decision making with application including aerospace, automotive, electronics and physics \cite{JONES202036, article, Journal, 6db924dfeff44d159ab577c1aefed6ef}. Traditional numerical simulations often rely on high-fidelity solvers, which are generally accurate, but expensive \cite{alma991043311449403276, sep-simulations-science, https://doi.org/10.1111/j.1467-8667.1989.tb00025.x}. This has historically motivated the development of reduced-order-models (ROM) which, at the cost of a drop in accuracy, can be much faster than high-fidelity full-order models (FOM). 

Many ROM methods rely on projecting FOM data into a lower dimensional space, where the physical system's governing equations are simpler to solve numerically \cite{doi:10.1146/annurev.fl.25.010193.002543, rbm, Safonov1988ASM}. Linear projection methods have been used with great success, but they often struggle with advection-dominated fluid flow problems \cite{https://doi.org/10.48550/arxiv.2009.11990, lasdi}. Thus, in recent years, non-linear approaches, using for instance neural networks, have gained significant popularity \cite{LEE2020108973, doi:10.1126/science.1127647, kutz_2017}. In their work, Champion et. al. \cite{doi:10.1073/pnas.1906995116} proposed a framework to identify governing dynamical systems in latent space of autoencoders. These identified governing sets of ordinary differential equations (ODEs), learned using SINDy algorithms \cite{doi:10.1073/pnas.1517384113, doi:10.1126/sciadv.1602614, Chen_2021, BONNEVILLE2022100115, STEPHANY2022360}, can be used to described the training data dynamics in a much simpler way. This work has been subsequently extended to reduced-order-modelling by Fries et. al., in a framework known as \textit{Latent Space Dynamics Identification} (LaSDI) \cite{lasdi}. 

In LaSDI, an autoencoder is trained on FOM data, and the set of ODEs governing the latent space for each corresponding FOM datapoint is identified using SINDy. The ODE coefficients can then be interpolated with respect to the FOM parameters, and used to predict sets of ODEs associated with any new test parameters. ROM predictions can then easily be made by integrating the ODEs numerically and feeding the dynamics into the decoder. This work was later extended in \textit{greedy-Latent Space Dynamics Identification} (gLaSDI) \cite{glasdi} by introducing an active learning framework and a joint training of the autoencoder and the latent space ODEs. During the training, additional data is added by feeding the model's prediction into the PDE residual. The simulation parameter yielding the largest residual error is picked as a sampling point, for which a new FOM run is performed and the resulting data added to the training set. This strategy allows for collecting data where it is the most needed, while minimizing the number of FOM runs. 

While robust and accurate, gLaSDI requires to have access to the PDE residual for sampling new data (\textit{intrusive} ROM). This can be cumbersome and expensive to evaluate, especially when dealing with multi-scale and/or multi-physics problems, and sophisticated numerical solvers. In this paper, we introduced \textit{Gaussian Process Latent Space Dynamics Identification} (GPLaSDI), a new LaSDI framework with \textit{non-intrusive} greedy sampling (no residual evaluation required). We propose to use Gaussian Processes (GP) to interpolate the sets of ODE coefficients, instead of deterministic methods like in LaSDI and gLaSDI \cite{lasdi, glasdi}. This allows, for new test parameters, to propagate the uncertainty over the ODE coefficients to the latent space dynamics, and then to the decoder output. Our approach has two major advantages: first, we can quantify the uncertainty over ROM predictions. Second, we can use that uncertainty to pick new FOM sampling datapoints, without relying on the PDE residual.

\section{Gaussian Process Latent Space Dynamics Identification}
We consider physical systems described by time-dependant PDEs of the following form:
\begin{equation}
\label{pde}
\frac{\partial \mathbf{u}}{\partial t}=\mathbf{f}(\mathbf{u},t,x\,|\,\pmb{\mu})\hspace{0.35in}(t,x)\in[0,t_\text{max}]\times\Omega\hspace{0.35in}\mathbf{u}(t=0,x\,|\,\pmb{\mu})=\mathbf{u_0}(x\,|\,\pmb{\mu})
\end{equation}
The vector of physical quantities $\textbf{u}$ is defined over a time interval $[0,t_\text{max}]$ and spatial domain $\Omega$. The PDE and its initial condition $\textbf{u}_0$ are parameterized by a parameter vector $\pmb{\mu}\in\mathcal{D}$. We assume that equation \ref{pde} can be solved with a high-fidelity FOM simulation, and we denote $\mathbf{U}^{(i)}=[\mathbf{u}_0^{(i)},\dots,\mathbf{u}_{N_t}^{(i)}]^\top\in\mathbb{R}^{(N_t+1)\times N_u}$ the matrix representing a numerical solution for parameter $\pmb{\mu}^{(i)}\in\mathcal{D}$, computed over $N_t$ time steps and $N_u$ spatial nodes. The collection of $N_\mu$ FOM data points is written $\mathbf{U}=[\mathbf{U}^{(1)},\dots,\mathbf{U}^{(N_\mu)}]$ 

We train an autoencoder on $\mathbf{U}$, and denote $\mathbf{Z}=\phi_e(\mathbf{U}|\mathbf{\theta}_\text{enc})\in\mathbb{R}^{N_\mu\times(N_t+1)\times N_z}$ the encoder output. $N_z$ is the latent space dimension, chosen arbitrarily (but such that $N_z\ll N_u$). The decoder output is $\mathbf{\hat{U}}=\phi_d(\mathbf{Z}|\mathbf{\theta}_\text{dec})$, and the training reconstruction loss is a standard mean-squared-error, $\mathcal{L}_{AE}(\mathbf{\theta}_\text{enc}, \mathbf{\theta}_\text{dec})=|\!|\mathbf{U}-\mathbf{\hat{U}}|\!|_2^2$. 

The architecture of the autoencoder is chosen such that the time dimension is unchanged. As a result, the latent space can be interpreted as a dynamical system, described by $N_z$ abstract variables. Hence, the encoder effectively transforms data described by a PDE into data described by an (unknown) system of ODEs, written as:
\begin{equation}
    \displaystyle\mathbf{\dot{Z}}^{(i)}=\psi_{DI}(\mathbf{Z}^{(i)}|\pmb{\mu}^{(i)})
\end{equation}
$\mathbf{\dot{Z}}^{(i)}$ refers to the time derivative of the set of latent space variables for data point $i\in[\![1,N_\mu]\!]$. For each training parameters $\pmb{\mu}^{(i)}$, we use an algorithm known as SINDy (\textit{Sparse Identification of Non-Linear Dynamics}) \cite{doi:10.1073/pnas.1517384113} to learn a suitable set of explicit governing ODEs. The key idea of SINDy is to approximate $\psi_{DI}$ as a linear combination between a library of potential candidate terms involved in the ODEs, $\mathbf{\Theta}(\mathbf{Z}^{(i)})$, and a matrix of coefficients $\mathbf{\Xi}^{(i)}$:
\begin{equation}
    \mathbf{\dot{Z}}^{(i)}\approx\mathbf{\Theta}(\mathbf{Z}^{(i)})\cdot\mathbf{\Xi}^{(i)\top}=\mathbf{\dot{\hat{Z}}}^{(i)}
\end{equation}
The library $\mathbf{\Theta}$ can include linear and non linear candidate terms, function of each latent space variable. The choice of terms is arbitrary, a broader variety of terms may capture the latent space dynamics more accurately, but also yield more complex sets of ODEs. In this paper, we restrict the library to linear terms only. The coefficient matrices are learned by minimizing the standard mean-squared-error, $\mathcal{L}_\text{SINDy}(\mathbf{\Xi})=|\!|\mathbf{\dot{Z}}-\mathbf{\dot{\hat{Z}}}|\!|_2^2$, with $\mathbf{\Xi}=[\mathbf{\Xi}^{(1)},\dots,\mathbf{\Xi}^{(N_\mu)}]\in\mathbb{R}^{N_\mu\times N_z\times N_l}$ ($N_l$ is the number of candidate terms in $\mathbf{\Theta}$). To ensure the learning of robust coefficients and sufficiently well conditioned sets of ODEs, the autoencoder and the SINDy coefficients are trained altogether. 

\begin{figure}
  \centering
  \includegraphics[width=\textwidth]{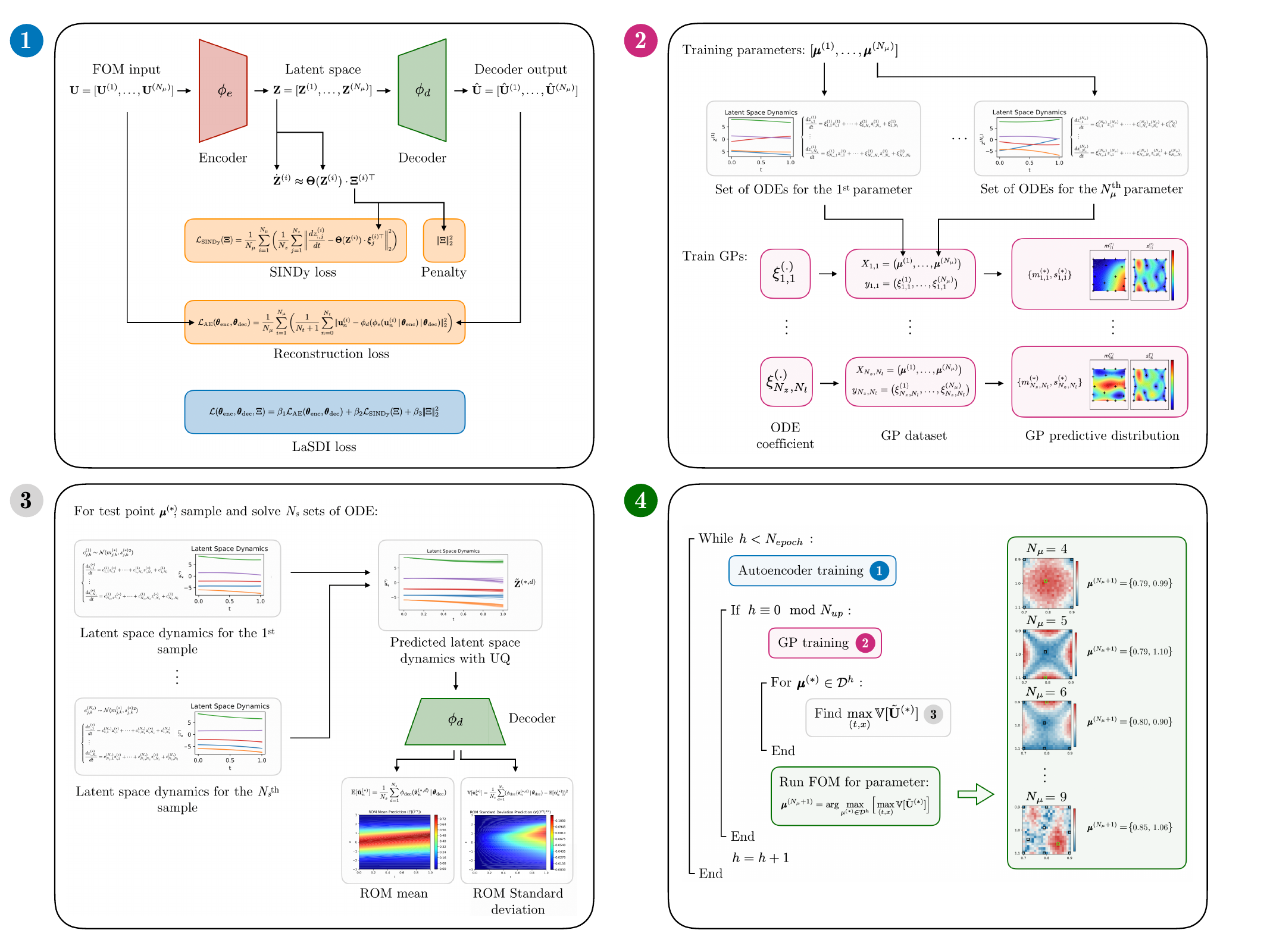}
  \caption{\label{framework} GPLaSDI framework. (1) Autoencoder/SINDy joint-training. (2) GP interpolation. (3) ROM prediction with uncertainty quantification. (4) Active learning strategy.}
\end{figure}

In order to make ROM predictions for a new test parameter $\pmb{\mu}^{(*)}\in\mathcal{D}$, we need to solve the corresponding set of ODEs, with coefficients $\mathbf{\Xi}^{(*)}$. In LaSDI \cite{lasdi} and gLaSDI \cite{glasdi}, $\mathbf{\Xi}^{(*)}$ is estimated through deterministic interpolation of $\mathbf{\Xi}$. In this paper, we propose to use Gaussian Processes (GPs). This has two main advantages: first, GPs are less prone to overfitting, which may provide a more accurate interpolation of the latent space governing dynamics. Second, GPs provide confidence intervals over their predictions. As a result, we can train GPs such that $\mathcal{GP}_\mathbf{\Xi}:\pmb{\mu}^{(*)}\mapsto\{m(\mathbf{\Xi}^{(*)}),s(\mathbf{\Xi}^{(*)})\}$, where $m$ and $s$ is the predictive mean and standard deviation of $\mathbf{\Xi}^{(*)}$, respectively. Note that in practice, we train $N_z\times N_l$ GPs, one for each ODE coefficient. We can than sample sets of ODE coefficients from the predictive Gaussian distribution, $\mathbf{\Xi}^{(*,d)}\sim\mathcal{N}(m(\mathbf{\Xi}^{(*)}),s(\mathbf{\Xi}^{(*)})^2)$, and integrate each set of ODEs, $\mathbf{\dot{Z}}^{(*,d)}=\mathbf{\Theta}(\mathbf{Z}^{(*,d)})\cdot\mathbf{\Xi}^{(*,d)\top}$, where $d\in[\![1,N_s]\!]$ and $N_s$ is the number of samples. Each sample set is solved using a numerical integrator (e.g. Forward Euler), and using initial condition $\mathbf{z}_0^{(*)}=\phi_e(\mathbf{u}_0^{(*)}|\mathbf{\theta}_\text{enc})$. Finally, we feed each latent space sample dynamics, $\mathbf{\Tilde{Z}}^{(*,d)}$ into the decoder to obtain a sample set of ROM predictions. We denote $\mathbb{E}[\mathbf{\Tilde{U}}^{(*)}]$ and $\mathbb{V}[\mathbf{\Tilde{U}}^{(*)}]$ the ROM mean prediction and variance, respectively. 

During the training phase, we adopt a variance-based greedy sampling strategy for active learning. We first train the autoencoder and the SINDy coefficients during $N_{up}$ epochs. Then, GPs are trained over the current set of learned ODE coefficients, $\mathbf{\Xi}$. For a finite number of test point $\pmb{\mu}^{(*)}\in\mathcal{D}^h\subset\mathcal{D}$ (where $\mathcal{D}^h$ is a discretized subset of the parameter space), we make ROM predictions and perform a new FOM simulation for the test parameter that yielded the highest ROM variance:
\begin{equation}
    \pmb{\mu}^{(N_\mu+1)}=\arg\max_{\pmb{\mu}^{(*)}\in\mathcal{D}^h}\big[\max_{(t,x)}\mathbb{V}[\mathbf{\Tilde{U}}^{(*)}]\big]
\end{equation}
The new data point, $\mathbf{U}^{(N_\mu+1)}$, is added to the rest of the training set and the training is resumed until the (preset) maximum number of epochs is reached. Figure \ref{framework} summarizes the GPLaSDI algorithm. 

\section{Results and Discussion}

We demonstrate the performance of GPLaSDI on a rising bubble problem. A hot fluid bubble is immersed in a colder fluid, causing the bubble to rise-up in a mushroom-like convective pattern. The solved PDE are the coupled compressible unviscid Navier-Stokes, advection-diffusion and energy conservation equations. In this example, implementing the residual would typically be lengthy and cumbersome, showcasing the interest of GPLaSDI.

We consider 2 simulation parameters: the initial bubble temperature, $\theta_c$, and its initial radius, $R_c$ ($\pmb{\mu}=\{\theta_c,R_c\}$). We employ \texttt{HyPar} \cite{HyPar, ghoshconstaAIAAJ2016}, a finite-difference PDE solver to perform high fidelity FOM simulations, and the numerical scheme used is described in \cite{jiangshu}. We consider a fully connected autoencoder, with softplus activation, a 10100-1000–200–50–20–5 architecture for the encoder, and a symmetric architecture for the decoder ($N_z=5$). We use Adam \cite{adam}, with learning rate $10^{-4}$ over $6.8\cdot10^5$ epochs, and consider a $0.25$ weighing factor for the SINDy loss. We use only linear terms in the SINDy library, and consider $N_s=20$ GP samples. We use GPs with RBF kernels, and a greedy sampling rate every $4\cdot10^4$ epochs. For testing purposes, the parameter space is discretized over a $21\times21$ grid ($\mathcal{D}^h$). As for the error metric, we consider the maximum relative error:

\begin{equation}
    e(\mathbf{U}^{(*)}, \mathbb{E}[\mathbf{\Tilde{U}}^{(*)}])=\max(|\!|\mathbb{E}[\mathbf{\Tilde{U}}^{(*)}]-\mathbf{U}^{(*)}|\!|_2/|\!|\mathbf{U}^{(*)}|\!|_2)
\end{equation}

\begin{figure}
  \centering
  \includegraphics[width=\textwidth]{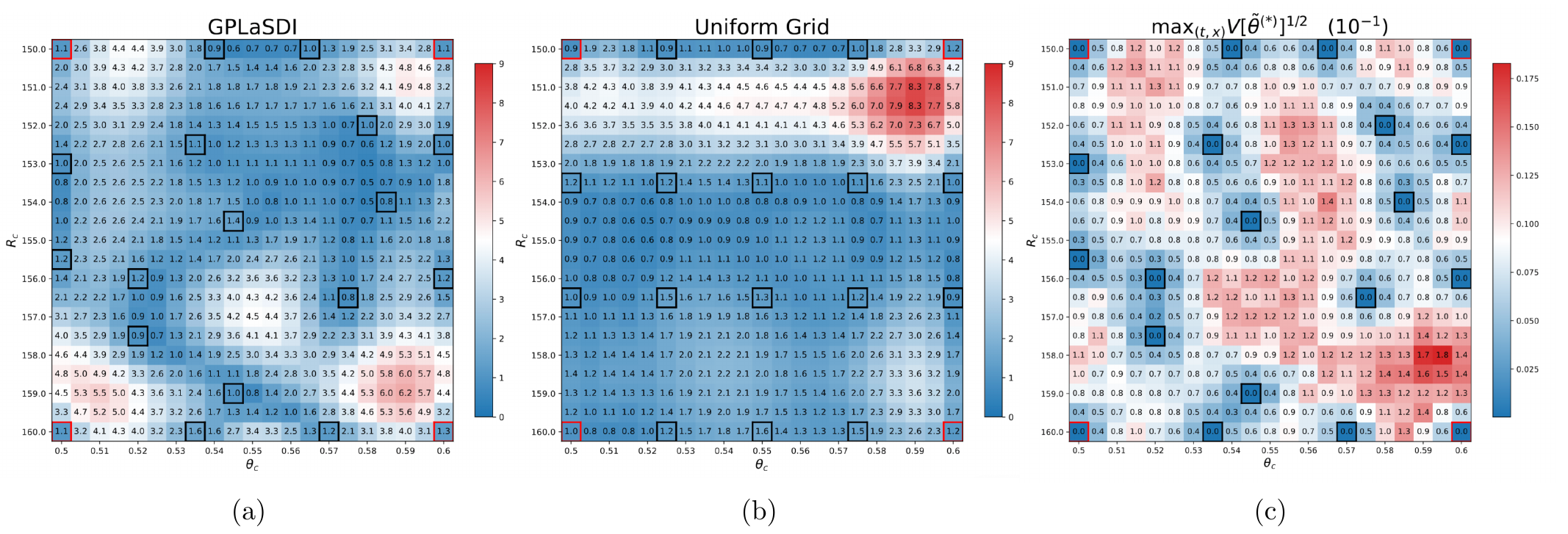}
  \caption{\label{risingbulle} (a) GPLaSDI maximum relative error (in percent). (b) LaSDI with uniform grid sampling maximum relative error. (c) GPLaSDI maximum predictive standard deviation. The 4 corner red boxes represents the parameters used in the initial training dataset. The black boxes represent the parameters sampled during training.}
\end{figure}

For baseline comparison, we compare the performance of GPLaSDI with uniform grid sampling. Figure \ref{risingbulle}.a shows the maximum relative error across the test parameter space $\mathcal{D}^h$ with GPLaSDI, and figure \ref{risingbulle}.b shows the results with uniform grid. With GPLaSDI, the worst maximum relative error is $6.2\%$, and the largest errors occur for parameter values located towards the bottom right corner of the
parameter space. GPLaSDI outperforms the baseline, which exhibits higher errors for smaller values of $R_c$ ($R_c<153$), with the worst maximum
relative error reaching $8.3\%$. The model is capable of efficiently sampling data points for which the uncertainty if the highest, and ultimately reduce the prediction error. Figure \ref{risingbulle}.c shows GPLaSDI maximum predictive standard deviation. It correlates reasonably well with the relative error, confirming that uncertainty-based sampling is effective in reducing model error. 

During 20 test runs, the FOM requires an average wall clock run-time of 89.1 seconds when utilizing 16 cores, and 1246.8 seconds when using a single core. On the other hand, the ROM model achieves an average run-time of $1.25\cdot10^{-2}$ seconds, resulting in an impressive average speed-up of $99,744\times$. Note that this speed up is obtained using only the predictive mean of the GPs for estimating the latent space set of ODEs (i.e. $N_s=1$). All the code to reproduce the results in this paper can be found at \href{https://github.com/LLNL/GPLaSDI}{\texttt{github.com/LLNL/GPLaSDI}}.

\section{Broader Impact}

We have introduced GPLaSDI, a fully data-driven non-intrusive ROM algorithm, with an uncertainty-based active learning approach. GPLaSDI is applicable to any type of physics or scientific phenomenon, does not requires direct access of even knowledge of the physical governing equations, and is capable of quantifying its own prediction uncertainty, while achieving remarkable speed-up.

\begin{ack}
This research was conducted at Lawrence Livermore National Laboratory and received support from the LDRD program under project number 21-SI-006. Y. Choi was supported for this work by the U.S. Department of Energy, Office of Science, Office of Advanced Scientific Computing Research, as part of the CHaRMNET Mathematical Multifaceted Integrated Capability Center (MMICC) program, under Award Number DE-SC0023164. Lawrence Livermore National Laboratory is operated by Lawrence Livermore National Security, LLC, for the U.S. Department of Energy, National Nuclear Security Administration under Contract DE-AC52-07NA27344. IM release: LLNL-CONF-855004
\end{ack}

\bibliography{references}

\end{document}